# Photonic antiferromagnetic topological insulator with a single surface Dirac cone

*Fujia Chen[1,2,3,6], Ning Han[1,2,3,6], Songyang Pu,[4,6] Rui Zhao[1,2,3], Li Zhang[1,2,3], Qiaolu Chen[1,2,3], Yuze Hu[1,2,3], Mingyu Tong[1,2,3], Wenhao Li[1,2,3], Junyao Wu[1,2,3], Yudong Ren[1,2,3], Xinrui Li [1,2,3], Wenyan Yin[1,\*], Hongsheng Chen[1,2,3,\*], Rui-Xing Zhang [4,5,\*], and Yihao Yang[1,2,3,\*]*

[1]*State Key Laboratory of Extreme Photonics and Instrumentation, ZJU-Hangzhou Global Scientific and Technological Innovation Center, Zhejiang University, Hangzhou 310027, China.*

[2]*International Joint Innovation Center, The Electromagnetics Academy at Zhejiang University, Zhejiang University, Haining 314400, China.*

[3] *Key Lab. of Advanced Micro/Nano Electronic Devices & Smart Systems of Zhejiang, Jinhua Institute of Zhejiang University, Zhejiang University, Jinhua 321099, China.*

[4] *Department of Physics and Astronomy, The University of Tennessee, Knoxville, Tennessee 37996, USA*

[5] *Department of Materials Science and Engineering, The University of Tennessee, Knoxville, Tennessee 37996, USA*

[6]*These authors contributed equally to this work.*

*\*e-mail:* (Y.Y.) *yangyihao@zju.edu.cn*; (W.Y.) *wyyin@zju.edu.cn* (H.C.) ;*hansomchen@zju.edu.cn*; (R.Z.) *ruixing@utk.edu*

**Abstract**

Antiferromagnetism, characterized by magnetic moments aligned in alternating directions with a vanished ensemble average, has garnered renewed interest for its potential applications in spintronics and axion dynamics. The synergy between antiferromagnetism and topology can lead to the emergence of an exotic topological phase unique to certain magnetic order, termed antiferromagnetic topological insulators (AF TIs). A hallmark signature of AF TIs is the presence of a single surface Dirac cone—a feature typically associated with strong three-dimensional (3D) topological insulators—only on certain symmetry-preserving crystal terminations. However, the direct observation of this phenomenon poses a significant challenge. Here, we have theoretically and experimentally discovered a 3D photonic AF TI hosting a single surface Dirac cone protected by the combined symmetry of time reversal and half-lattice translation. Conceptually, our setup can be viewed as a *z*-directional stack of two-dimensional Chern insulators, with adjacent layers oppositely magnetised to form a 3D type-A AF configuration. By measuring both bulk and surface states, we have directly observed the symmetry-protected gapless single-Dirac-cone surface state, which shows remarkable robustness against random magnetic disorders. Our work constitutes the first realisation of photonic AF TIs and photonic analogues of strong topological insulators, opening a new chapter for exploring novel topological photonic devices and phenomena that incorporate additional magnetic degrees of freedom.

**Introduction**

The interplay between topology and magnetism gives rise to many exotic quantum phenomena, such as the quantum anomalous Hall effect and quantum spin liquid[1-3], where the breaking of time-reversal symmetry (TRS) is crucial. Meanwhile, magnetic systems often feature built-in anti-unitary symmetries that can underpin novel topological structures unique to certain types of magnetic ordering[4]. A prominent example is three-dimensional (3D) antiferromagnetic topological insulators (AF TIs)[5], where magnetic moments align alternately, resulting in a vanished ensemble average. These AF TIs typically respect an emergent anti-unitary symmetry $S = \Theta T_{1/2}$, combining TRS $\Theta$ with a half-lattice translation $T_{1/2}$, which links two adjacent, oppositely aligned magnetic moments. Notably, the $S$ symmetry can stabilise a $\mathbb{Z}_2$ topological classification of electronic insulators, with the $\mathbb{Z}_2$ nontrivial phase exhibiting gapless single-Dirac-cone surface states—features commonly observed in TRS-invariant strong 3D TIs[6,7], but only on the $S$-preserving surfaces. Due to their profound underlying physics and potential applications in axion dynamics and spintronics[4,5,8-14], the realisation of AF TIs has been a decade-long endeavour, with van der Waals layered MnBi$_2$Te$_4$ identified as the first viable real-world candidate only recently[15,16]. Despite the significant experimental effort, the direct observation of the gapless surface Dirac cone of AF TIs has remained elusive, primarily due to the experimental challenges associated with directly probing the $S$-preserving crystal surfaces in MnBi$_2$Te$_4$.

On the other hand, fermions and bosons often adhere to distinct rules of symmetry and topology[17,18]. For example, the TRS operator for spinful fermions, like electrons, satisfies $\Theta_f^2 = -1$, whereas for bosons or spinless fermions, such as photons and magnons, it follows $\Theta_b^2 = 1$. Specifically, $\Theta_f$ enables the famous Kramers' theorem for electrons, which can support a TR-invariant strong 3D TI with an odd number of topological Dirac surface states. However, an equivalent bosonic counterpart does not exist. Unlike the TR-invariant TIs, the concept of AF TI applies to bosonic systems but appears fundamentally different[19]. This manifests in $S^2 = \Theta^2 e^{ik_D a_D}$, with $k_D$ and $a_D$ a crystal momentum and a lattice constant along $T_{1/2}$, respectively. The $k_D = \pi/a_D$ plane of a 3D bosonic system features an effective fermionic TRS $S^2 = -1$, thereby making both the Kramers' theorem and $\mathbb{Z}_2$ topology well-defined. Consequently, the bosonic AF TI features a single surface Dirac cone at $k_D = \pi/a_D$ instead of $k_D = 0$. To date, however, the bosonic AF topology has never been realised.

Here, we report the first experimental realisation of a 3D photonic AF TI featuring a single gapless $S$-protected Dirac surface state. Our setup is effectively configured as a $z$-directional stack of two-dimensional (2D) Chern insulators[1,20], with adjacent layers oppositely magnetised to establish a 3D type-A AF configuration[5]. The nontrivial $\mathbb{Z}_2$ bulk topological invariant is numerically confirmed by tracking the evolution of Wannier centres in first-principles simulations. By directly mapping out both bulk and surface band structures, we have unambiguously verified the existence of a single symmetry-protected surface Dirac cone at the $k_z = \pi/a_z$ plane, the defining characteristic of bosonic AF TIs. Our photonic AF TI can also be regarded as the first realisation of a 3D strong photonic TI in terms of the odd number of surface Dirac cones. Additionally, our experiments have demonstrated the remarkable resilience of the single-Dirac-cone surface state against random magnetic disorders that, on average, respect the $S$ symmetry. Our findings pave the way for further realising and engineering novel magnetic topological phenomena in photonic systems, which are inaccessible in electronic systems.

**Results**

We start with a realistic design of photonic AF TIs with a 3D hexagonal unit cell containing two magnetically-biased yttrium-iron-garnet (YIG) rods and two metallic plates perforated with holes (Fig. 1a,b). The holes, located at the corners of the hexagonal unit cell with two different diameters of $d_1$ and $d_2$, introduce interlayer coupling and control the in-plane rotation symmetry. Two permanent magnets are snugly placed on either side of YIG rods to magnetise them along the $z$ axis (see Method for the parameters of YIG rods). The magnetisation aligns in the same direction for YIG rods for the same layers but opposite for adjacent layers, exhibiting a 3D type-A "antiferromagnetic configuration". Under the Bloch-wave basis, $S$ symmetry in our design yields

$$S^2 = e^{ik_z h}, \qquad (1)$$

where $h$ is the lattice constant in the $z$ direction, and $k_z$ is the $z$ component of Bloch wavevectors. Therefore, we have $S^2 = -1$ at $k_z = \pi/h$ plane, a reminiscence of the spinful TR operation in fermionic systems. Consequently, the $k_z = \pi/h$ plane achieves an effective 2D class-AII system where (i) Kramer's degeneracy can be applied and (ii) a 2D $\mathbb{Z}_2$ topology is well-defined [19].

We first consider a photonic AF TI design respecting an in-plane $C_6$ symmetry, e.g., $d_1 = 5.4$ mm and $d_2 = 5.4$ mm. The simulated band structure of our design is shown in Fig. 1c, which exhibits a sizable band gap throughout the 3D Brillouin zone. Notably, the energy bands are doubly degenerate everywhere at the $k_z = \pi/h$ plane but not at $k_z = 0$ (see Fig. 1d), a direct outcome of the Kramers' theorem of $S$. In Fig. 1e, we further confirm that a pair of $S$-related orthogonal modes constitute a pseudospin degree of freedom. The class-AII nature allows us to define a 2D $\mathbb{Z}_2$ topological index[6], which can be numerically evaluated by counting the relative winding number of the spectral flow of the Berry phase using the Wilson loop method[21]. In Fig. 1f, we plot the calculated spectrum of the Berry phase of the two lowest bands at the $k_z = 0$ and $\pi/h$ planes, respectively (see Methods for numerical calculations). For $k_z = \pi/h$, the gapless flow of Berry phase from $-\pi$ to $\pi$ along $k_1$ (see inset in Fig. 1f) clearly indicates a nontrivial $\mathbb{Z}_2$ invariant, which ensures the existence of a single surface Dirac cone on the $S$-preserving surfaces. Interestingly, the Berry phase at the $k_z = 0$ plane shows a similar helical pattern despite the break-down of Kramers' theorem ($S^2 = 1$), which arises from an emergent fragile topology protected by $C_6$ symmetry[22] (see Supplementary Information for more detail).

Next, we experimentally fabricate the photonic AF TI. Figure 2a-c show the fabricated sample consisting of 12 unit cells or 24 sublayers in the $z$ direction. Constrained by the experiments, the single surface Dirac cone of the 3D photonic AF TI design in Fig. 1c is not significantly evident (see surface dispersion in Extended Data Fig. 1), which hinders its direct observation. To facilitate the experimental demonstration, we modify the hole diameters of the coupling layers, $d_1 = 6.4$ mm and $d_2 = 4.4$ mm, which reduces the system's in-plane symmetry from $C_6$ to $C_3$. The symmetry reduction lifts the band degeneracy at the $k_z = \pi/h$ plane except for TRS-invariant momenta **A** and **L** (see Extended Data Fig. 2). Notably, this structural update spoils the fragile topology at $k_z = 0$ while preserving the key $\mathbb{Z}_2$ topology of our AF TI. The dielectric foam plates with a relative permittivity of 1.05 and a loss tangent of 0.0005 are drilled with a triangular array of 3 mm diameter holes to fix the position of the YIG rods and for inserting the source and the probe. The coupling layers are machined from iron to avoid repulsive magnetic forces between adjacent magnets effectively, and these coupling layers have negligible influence on the intralayer magnetisations (see Methods for more details).

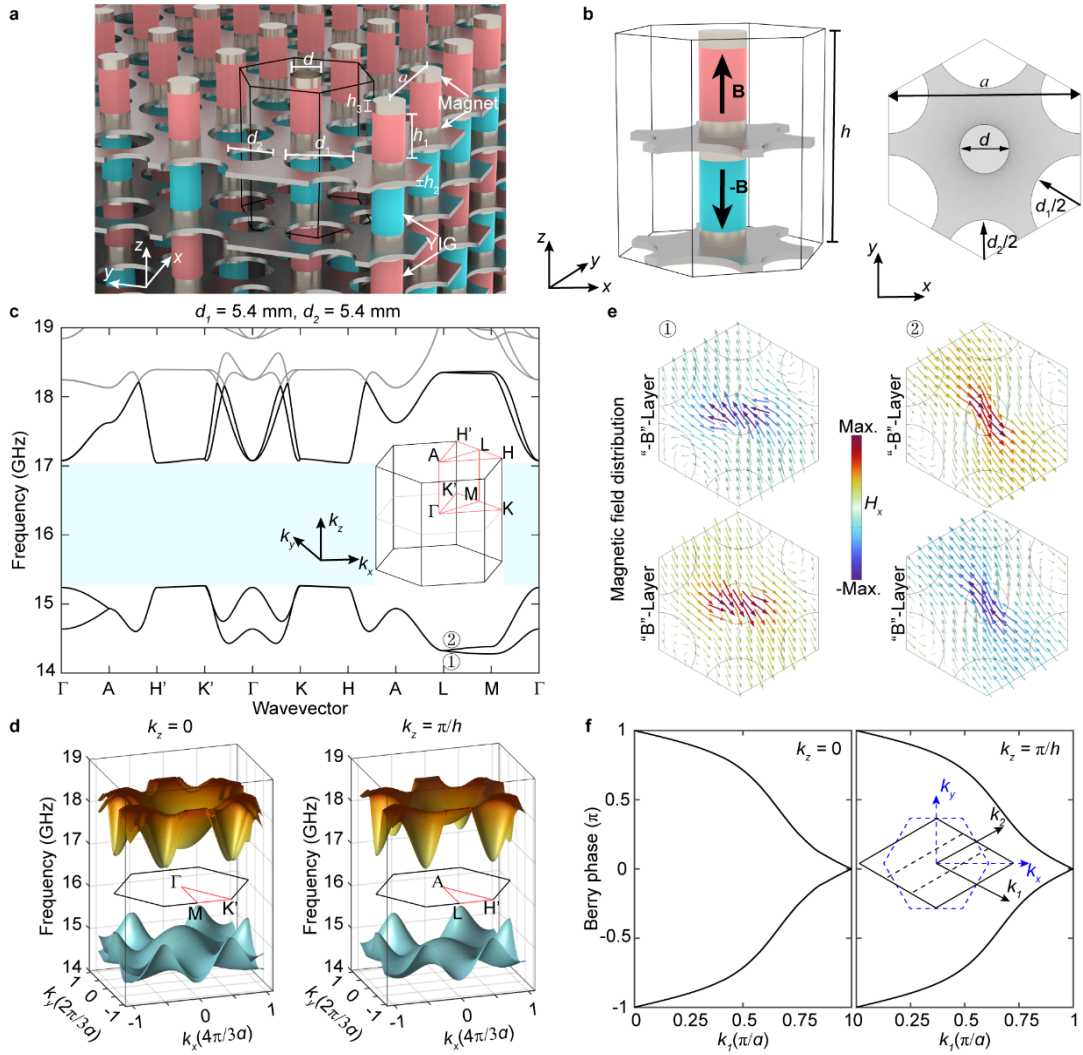

**Fig.1 | Design of a 3D photonic AF TI. (a, b)** Schematic diagram (**a**) and a unit cell (**b**) of the photonic AF TI. The photonic AF TI consists of stacked 2D photonic Chern insulators with alternating Chern numbers along the z direction, interacting with each other with elaborately designed coupling layers. YIG rods are arranged in a triangular lattice for each layer and magnetised with a pair of permanent magnets. The black arrows in (**b**) represent the opposite magnetisation direction for YIG rods of two adjacent layers. The structure parameters are $d = 3$ mm, $h_1 = 4.5$ mm, $h_2 = 0.5$ mm, $h_3 = 1$ mm, $d_1 = 6.4$ mm, and $d_2 = 4.4$ mm. The lattice constant in xy-plane is $a = 11.6$ mm, and the periodicity along the z direction is $h = 14$ mm. The background foam material has a relative permittivity of 1.05, and a loss tangent of 0.0005. (**c**) Simulated bulk band structures of the photonic AF TI along high-symmetry lines with $d_1 = 5.4$ mm, and $d_2 = 5.4$ mm. The transparent blue region represents the complete 3D topological bandgap. Inset: 3D bulk BZ of photonic AF TI. (**d**) 2D band structure at the momentum plane with $k_z = 0$ (left panel) and $\pi/h$ (right panel), respectively. (**e**) Simulated 2D cutting plane of the magnetic field distributions of the eigenstates at lower bands near **L**. The cutting planes are located in the middle of the YIG rods. (**f**) Calculated spectral flow of Berry phase of two bands below the bandgap for $k_z = 0$ (left panel), and $\pi/h$ (right panel) planes, respectively. Inset in (**f**) represents the original 2D triangular (dashed blue hexagon), and the deformed (solid black rhombus) BZs for calculating Berry phases.

Then, we perform experiments to characterise the bulk states of the photonic AF TI. To measure the bulk dispersion, a dipole source antenna is inserted from the bottom into the centre of the sample; a second dipole antenna, acting as the probe, is inserted into the bulk from the top to map out the complex electric field distributions hole by hole, in a (010) plane inside the sample 6.7 mm away from the source (see Fig. 2e). The frequency dependence of the transmittance in the bulk is shown in Fig. 2f. One can see a stop band extending from approximately 15.6 to 16.4 GHz, corresponding to a complete 3D bulk bandgap with a vanished density of states (DOS)[23]. The excited fields within the bandgap are strongly confined around the source, as shown in Fig. 2e. After applying the Fourier transform to the measured complex field distributions, we obtain the measured projected bulk band structure along high-symmetry lines of the projected surface BZ (see Fig. 2f). One can see over the bandgap, no visible states are excited. Moreover, the measured result closely matches the simulated counterpart.

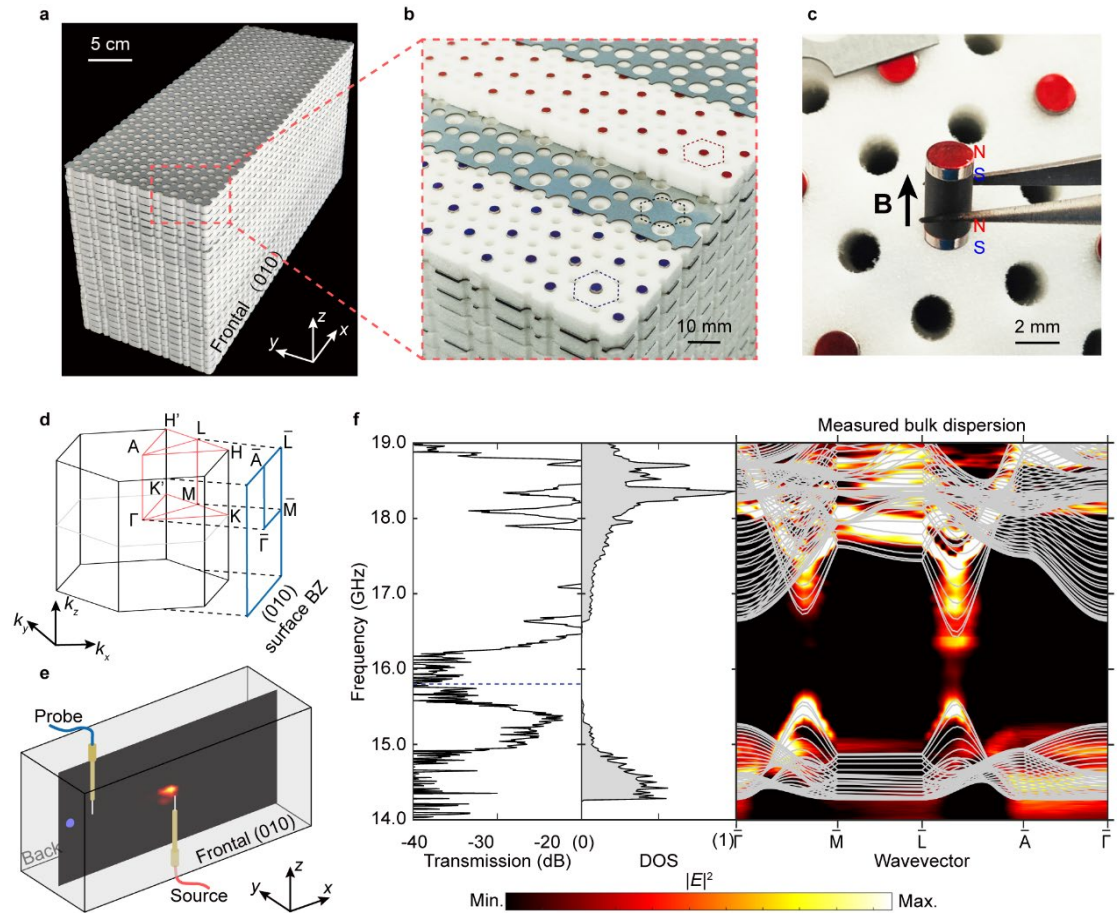

**Fig. 2 | Experimental setup and measured bulk dispersion.** (**a-c**) Photograph of the fabricated photonic AF TI. (**b, c**) Enlarged view of the sample and the YIG rod. In (**b**), the top three layers are moved for clarity. The YIG rods sandwiched between two permanent magnets are embedded in the dielectric foam. The face colours in (**b**) and (**c**) represent the N- (red) and S- (blue) poles of magnets. The black arrow in (**c**) represents the direction of the external magnetic field. The holes in the dielectric foam are used to insert the source and the probe. (**d**) 3D bulk BZ and its projection onto the (010) surface BZ. (**e**) Experimental setup and the measured electric intensity on (010) plane at 15.8 GHz. The source is positioned at the sample centre, while the probe sweeps the selected plane,

hole by hole, to map the complex electric field distributions. (**f**) Measured bulk transmission (left panel), calculated density of states (DOS, middle plane), and projected bulk band structure (right panel) of the photonic AF TI. The transmission is measured at the position marked by the blue sphere in (**d**). The vanished DOS indicates a complete bandgap. Grey curves in the right panel display the simulated projected bulk band structure. The colour scale below (**f**) measures the electric energy density $|E|^2$.

The 3D photonic AF TI hosts single-Dirac-cone surface states at *S*-preserving surfaces, e.g., the (010) surface. In Fig. 3a, we plot the simulated surface dispersion along high-symmetry lines of frontal (010) surface BZ (see Fig. 3b) and the 2D band structure near $\bar{L}$. The surface dispersion traverses the entire topological bandgap and exhibits a single Dirac cone at $\bar{L}$ with a Dirac frequency at 15.72 GHz. Such surface states are strongly localised near the sample surface (see Fig. 3c). As a comparison, we have also plotted the numerically calculated Berry phase along the same momentum-space trajectory as our surface plot, which is not only gapless but also homotopic to the surface band structure. This feature exemplifies the bulk-boundary correspondence principle. To measure the topological surface states, we cover the sample frontal (010) surface with a copper plate and other surfaces with microwave absorbers. A source antenna is inserted from the bottom into the centre of the sample frontal (010) surface, while the probe is inserted from the top to map the complex field distributions at the surface (see Fig. 3e). The measured surface dispersion, shown in Fig. 3f, exhibits clearly a single surface Dirac cone with the Dirac point around 15.72 GHz. The Dirac cone can be further confirmed from the measured iso-frequency contours in the 2D reciprocal space, shown in Fig. 3g. The experimental results closely match the numerically simulated counterparts. All these results provide direct experimental evidence of the single surface Dirac cone at $k_z = \pi/h$ on the *S*-invariant surfaces, a hallmark signature of the photonic AF TI.

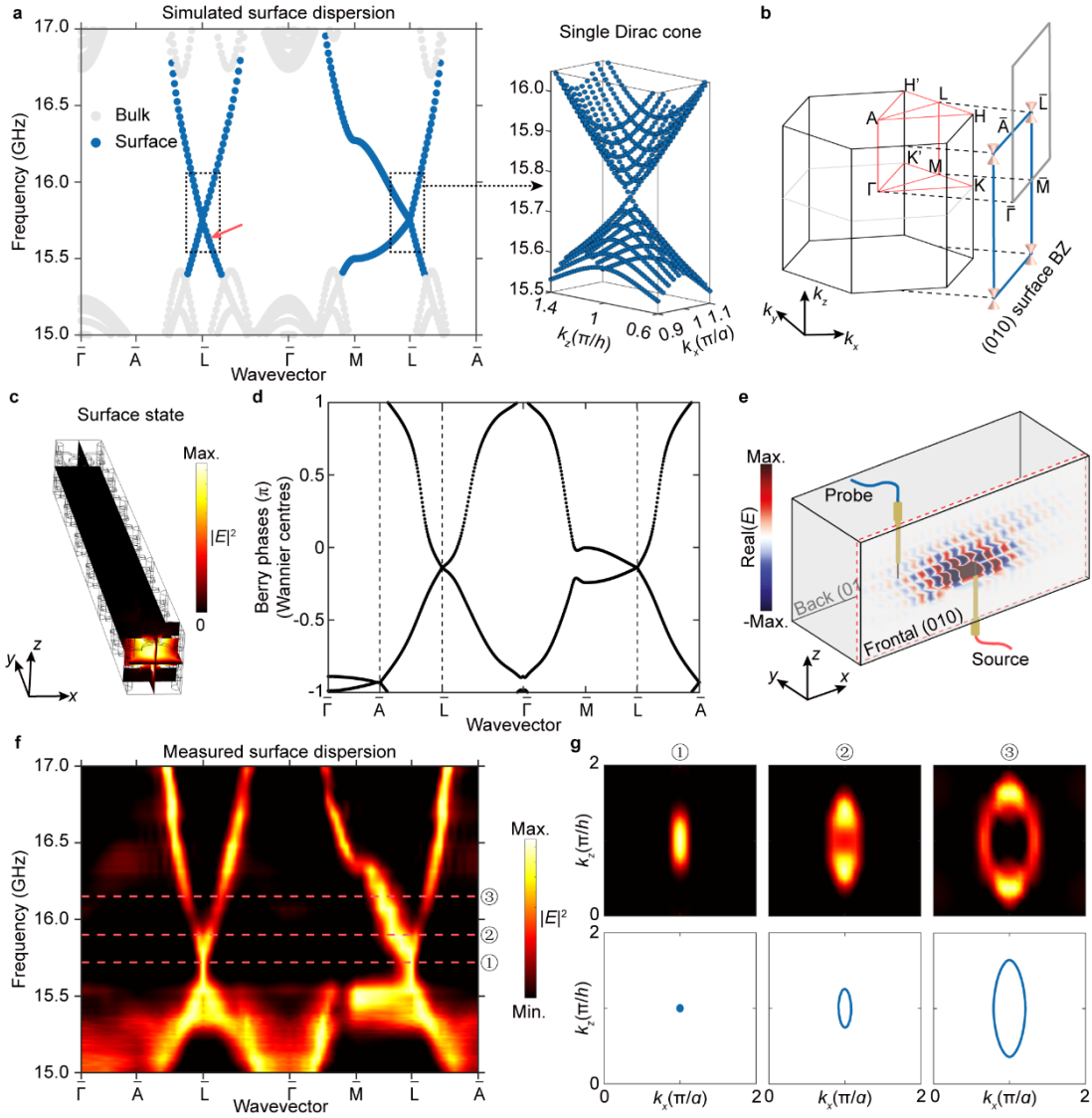

**Fig. 3 | Experimental observation of a single surface Dirac cone on the *S*-symmetry preserved surface.** (**a**) Simulated surface dispersion (left panel) along high-symmetry lines and 2D surface dispersion (right panel) around $\bar{L}$. Blue (Grey) circles represent surface (projected bulk) states. The simulated surface dispersion manifests a single surface Dirac cone located at $\bar{L}$. (**b**) 3D bulk BZ and its projection onto the (010) surface BZ with a single surface Dirac cone at the corner. The grey tetragon represents the surface BZ centered at $\bar{L}$. (**c**) Simulated field energy distribution of the surface state, corresponding to the pink arrow in (**a**). (**d**) Calculated spectral flow of Berry phases (or "hybrid Wannier centres") along high-symmetry lines in the surface BZ. The degeneracy at $\bar{L}$ indicates the existence of the single surface Dirac point. The evolution of bulk Berry phases is homotopic to the dispersion shown in (**b**). (**e**) Experimental setup and measured field distribution on sample frontal (010) surface at the Dirac point (15.72 GHz). (**f**) Measured surface dispersion along high-symmetry lines. (**g**) Measured (top panels) and simulated (bottom panels) iso-frequency contours of the single surface Dirac cone at different frequencies. The point-like contour in the first column represents the Dirac point, and the other elliptic contours indicate the anisotropy of the surface Dirac cone. The pink dashed lines in (**f**) label the frequencies of (**g**).

The symmetry-protected single-Dirac-cone surface states have been theoretically assumed to be resilient against random disorder of any type on the surface[24-26]. To demonstrate this point, we introduce magnetic disorder to the outermost unit cells by randomly flipping the original magnetisation direction as schematically shown in Fig. 4a. A source is placed at the right centre of the sample frontal (010) surface to launch the leftwards propagated surface states, then a field measurement as before is performed. Specifically, we altered the magnetization direction of 5%, 10%, and 15% of the total magnets in separate experiments (see Fig. 4d-f). These customisable magnetic disorders can be challenging to achieve and manipulate in electronic systems, showing the advantage of our photonic system. In Fig. 4b, the transmissions measured at different disorder density exhibit a similar frequency-dependent pattern and are significantly greater than the bulk transmission within the topological bandgap, all of which further resemble the disorder-free case. This suggests the gapless nature of Dirac surface state remains in the presence of disorders. For four disorder configurations, the field patterns, at 15.85 GHz, also distribute in the same way as the traveling wave (Fig. 4c,g-i), which demonstrates again that the surface states remain delocalised even in the presence of strong disorder. After Fourier transformation to momentum space, we observe that the momenta of surface states only distribute to the right of the isofrequency contour, indicating that only leftwards propagated states are excited and magnetic disorders do not lead to backscattering.

Although these magnetic disorders, in our case, break $S$ symmetry locally, their ensemble average does not because of the random distribution[24]. We now offer an intuitive interpretation from the percolation picture. Note that the surface-state photons propagate in the clean region of 2D surface like a massless Dirac particle. The Dirac particle acquires a positive or negative mass term if the $S$ symmetry is broken locally. Since the opposite Dirac mass term must occur in the adjacent region on the surface, a chiral edge mode exists in the domain wall between two regions, similar to the one-way edge states in photonic Chern insulators[26,27]. When the disorders vanish on average, our system is equivalent to the critical state of a quantum Hall plateau transition[28], where the carriers are necessarily delocalised.

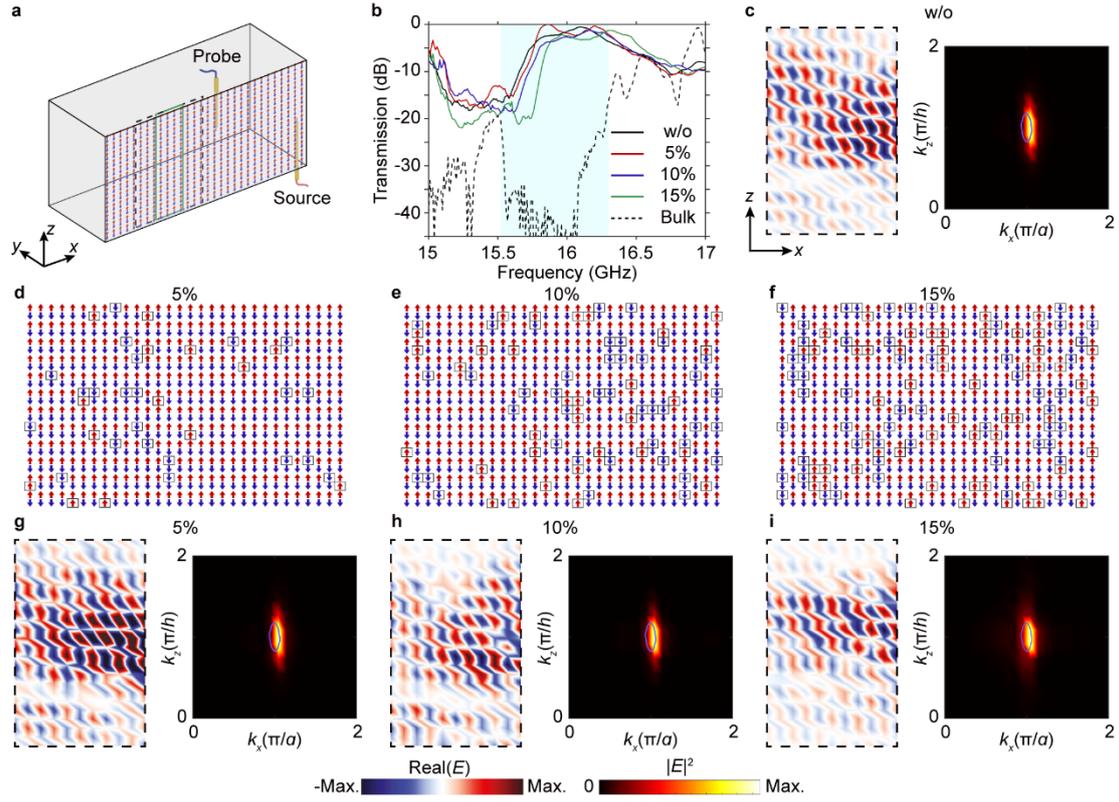

**Fig. 4 | Robust single-Dirac-cone topological surface states in the presence of random magnetic disorder.** (**a**) Experimental setup. The source is positioned at the right of the surface. Red and blue arrows represent the opposite magnetisation direction of YIG near the sample frontal (010) surface. The green box indicates the position of measuring the surface-state transmission. The black dashed box indicates the region of measuring field distribution. (**b**) Measured transmission for surface states with 0% (black line), 5% (red line), 10% (blue line), and 15% (green line) random magnetisation direction reversal and for bulk states (black dashed line). The region highlighted in transparent blue represents the bulk topological bandgap. (**c**) Measured field distribution (left panel) and corresponding isofrequency contour (right panel) at 15.85 GHz in the case of without disorder. (**d-f**) Experimental settings for disorder strength of 5% (**d**), 10% (**e**), and 15% (**f**). (**g-i**) Measured field distribution (left panel) and corresponding isofrequency contour (right panel) at 15.85 GHz in the cases with disorder strength of 5% (**g**), 10% (**h**), and 15% (**i**), respectively.

We have successfully developed both theoretical and experimental frameworks to establish a 3D photonic AF TI. As a hallmark signature of AF TIs, the gapless single Dirac cone on $S$-preserving surfaces has been directly visualised, which has been notably challenging to accomplish in electronic systems. Furthermore, we have experimentally demonstrated the exceptional robustness of the single-Dirac-cone surface states against strong random magnetic disorders that preserves $S$ on average. Our work thus expands the scope of AF TIs from fermions to bosons. Many intriguing phenomena of strong 3D topological insulators are associated with the odd number of surface Dirac cones, such as the image magnetic monopole[29], which may find their bosonic counterparts within our photonic platform. Intriguingly, by gapping out the single surface Dirac cone, it may be possible to realize 3D axion insulators[4,5,8-10]. Moreover, our photonic AF TI paves the way for probing a variety of 3D photonic topologies that emerge from the complex interplay between magnetic order and symmetry across the 1,651 known magnetic groups[30]. Finally, our experience with the photonic AF TI can also be applied to other bosonic systems, such as phonons, magnons, and polaritons.

**Methods**

**Numerical simulations.** A finite-element method solver (COMSOL Multiphysics RF Module) is used to perform all numerical simulations in this work. To calculate the 3D bulk dispersion and projected bulk dispersion of 3D photonic AF TI, periodic boundary conditions are applied in all directions of the unit cell. For the surface dispersion in Fig. 3(b), we consider a supercell consisting of 13 unit cells; Periodic boundary conditions are imposed in the $x$ and $z$ directions, with perfect electric conductor (PEC) boundary condition in the y-direction. In all simulations, iron coupling layers and magnets are modelled as PECs (see more details in Supplementary Information).

The relative permeability tensor of magnetised YIG has the form of

$$[\mu_r] = \begin{bmatrix} \mu_m & \pm j\mu_k & 0 \\ \mp j\mu_k & \mu_m & 0 \\ 0 & 0 & 1 \end{bmatrix}, \quad (1)$$

where $\mu_m = 1 + \omega_m(\omega_0 + i\alpha\omega)/[(\omega_0 + i\alpha\omega)^2 - \omega^2]$, $\mu_k = \omega_m\omega/[(\omega_0 + i\alpha\omega)^2 - \omega^2]$, $\omega_m = \mu_0\gamma M_s$, $\omega_0 = \mu_0\gamma H_0$, and $\gamma H_0$ is the external magnetic field (about 0.24 Tesla) along the z-direction, $\gamma = 1.759 \times 10^{11}$ C/kg is the gyromagnetic ratio, $\alpha = 0.0088$ is the damping coefficient, and $\omega$ is the operating frequency. The dispersions of the permeability tensor elements $\mu_m$ and $\mu_k$ under 0.24 Tesla external magnetic field are shown in Fig. S1 of Supplementary Information, respectively. Because the resonance frequency of the permeability is far from topological bandgap, to facilitate numerical simulations we ignored weak dispersion in permeability within the frequency range of interest.

**Calculation of Berry phases.** The Berry phases are calculated through the linking matrices $M_{mn}^{k,k+\Delta k} = \langle u_{mk} | u_{n(k+\Delta k)} \rangle$ of the Block wavefunctions $u_{mk}$ between the neighbouring two $k$ points of degenerate bands. We multiply the linking matrices to be the Wilson loop $W(k_\perp) = \Pi M^{k_\parallel, k_\parallel + \Delta k_\parallel}$ along the closed loop in the BZ-a parallel transport cycle. The Wilson loop eigenvalues are $\lambda_n(k_\perp)$, and the Berry phases are $\phi_n(k_\perp) = \text{Im}[\log \lambda_n(k_\perp)]$. The hybrid Wannier centres $\bar{z}_n(k_\perp) = \frac{c}{2\pi}\phi_n(k_\perp)$ [21]. Here, $c$ is the real-space period in the direction of the surface, which is $\frac{2a}{\sqrt{3}}$ in our hexagonal unit cell for (010) surface.

**Qualitative tight-binding models.** The 3D photonic AF TI proposed in this work can be qualitatively described by a 3D tight-binding model. This tight-binding model is a layer-by-layer stacking of the 2D Haldane model along z direction with interlayer couplings, but with opposite magnetic flux Φ between two adjacent layers, as shown in Extended Data Fig. 3a. The unit cell contains four different sublattice sites, the A1 (orange) and B1 (purple) on bottom layer (light blue hexagon), the A2 (orange) and B2 (purple) on top layer (light red hexagon). A1 (A2) and B1 (B2) sublattice sites possess on-site potential of $M$ and $-M$, respectively. The intralayer couplings contain the nearest-neighbor (NN) coupling $t_1$ (solid black lines) and next-nearest-neighbor (NNN) couplings $t_2 e^{\pm i\Phi}$ (solid orange and purple lines, '$\pm$' represents two adjacent layers having opposite $\Phi$). The interlayer couplings are $t_a$ (dashed orange lines) for A1 and A2 sublattice sites and $t_b$ (dashed purple lines) for B1 and B2 sublattice sites, respectively. Thus, the corresponding Bloch Hamiltonian of this model is given by a 4 × 4 matrix as

$$H = \begin{pmatrix} H_+ & T \\ T^\dagger & H_- \end{pmatrix}, \quad (2)$$

where

$$\begin{aligned}H_\pm =\ & M\sigma_3 + t_1\left(\cos k_1 + \cos k_2 + 1\right)\sigma_1 + t_1\left(\sin k_1 + \sin k_2\right)\sigma_2 \\ & + 2t_2 \cos\Phi\left(\cos k_1 + \cos k_2 + \cos(k_1 - k_2)\right)\sigma_0 + \\ & \pm 2t_2 \sin\Phi\left(\sin k_1 + \sin k_2 + \sin(k_1 - k_2)\right)\sigma_0\end{aligned} \quad (3)$$

and $\sigma_i$ (i = 1, 2, 3) are Pauli matrices, and $\sigma_0$ is a 2 × 2 identity matrix. Here, we define $k_{1,2} = \boldsymbol{k} \cdot \boldsymbol{a}_{1,2}$ where $a_1 = a(3/2, \sqrt{3}/2)$, $a_2 = a(3/2, -\sqrt{3}/2)$ are the in-plane lattice vectors of the honeycomb lattice. The interlayer coupling T can be expressed as

$$T = \begin{pmatrix} t_a\left(1 + e^{ik_z h/2}\right) & 0 \\ 0 & t_b\left(1 + e^{ik_z h/2}\right) \end{pmatrix}, \quad (4)$$

where $h$ is the lattice constant along the $z$ direction.

Tight-binding Hamiltonian in Eq. (4) preserves $S$ symmetry, that is $SH(\boldsymbol{k})S^{-1} = H(-\boldsymbol{k})$ (see Supplementary Information for more symmetry analysis). To demonstrate the effectiveness of this Hamiltonian, in Extended Data Fig. 3, we study its bulk and surface properties in the case of preserving ($M = 0$, $t_a = 1$, and $t_b = 1$) and breaking ($M = 1$, $t_a = 1.9$, and $t_b = 0.1$) in-plane $C_6$ symmetry. We can observe that for both cases, the tight-binding model reveals the similar phenomena as the realistic design.

**Materials and experimental setups.** In the experiment, we adopt commercially available yttrium-iron-garnet (YIG) ferrites and permanent magnets to break the time-reversal symmetry. The radius and height of YIG ferrites are 1.5 mm and 4.5 mm, respectively. The YIG ferrites have a saturation magnetisation $M_s$ = 1780 Gauss. Its permittivity and permeability are about $\varepsilon_r$ = 14.3 + 0.003$i$ and $\mu_r$ = 1, which are essentially constant microwave frequencies. The permanent magnets are electroplated by a nickel with a thickness of 0.002 mm. A pair of magnets provides an overall uniform external magnetic field of about 0.24 Tesla to magnetise the YIG rods. The iron plates are perforated with air holes precisely using a Computer Numerical Control (CNC) machine. To fix the position of the sandwiched YIG rods and magnets, we adopt the perforated dielectric foam with a relative permittivity 1.05, and a loss tangent 0.0005.

In the measurements, two microwave dipole antennas function as source and probe are connected to a vector network analyser (Ceyear 3672B). By inserting the probe into the air holes one by one and scanning along the z-direction in the half-lattice steps, we can map the complex electric field distribution on the surfaces or within the bulk of the experimental sample. In the bulk dispersion measurement, we wrapped all sides of the sample with microwave absorbers. In surface dispersion measurement, we cover the sample's frontal (010) surface with metallic claddings and the sample's right and left (100) surfaces with absorbers. 2D Fourier transformation is performed to the measured complex electric field distributions at each frequency to obtain the projected bulk and surface dispersions.

**Data availability**

The data in this study are available.

**Acknowledgements**

The work at Zhejiang University was sponsored by the Key Research and Development Program of the Ministry of Science and Technology under Grants No. 2022YFA1405200 (Y. Y.), 2022YFA1404704 (H. C.), 2022YFA1404902 (H. C.), and 2022YFA1404900 (Y. Y. ), the National Natural Science Foundation of China (NNSFC) under Grants No. 62175215 (Y. Y.), No. 62071418 (W. Y.), and No. 61975176 (H. C.), the Fundamental Research Funds for the Central Universities (2021FZZX001-19) (Y. Y.), the Excellent Young Scientists Fund Program (Overseas) of China (Y. Y.), and the Key Research and Development Program of Zhejiang Province under Grant No. 2022C01036 (H. C.). S.P. and R.X.Z. are supported by a start-up fund at the University of Tennessee.

**Author contributions**

Y.Y. and R.X.Z. initiated the idea. F.C., S.P., R.X.Z., and YY provided theoretical explanations. F.C. performed the numerical simulations. F.C. and Y.Y. designed the experiment. F.C., R.Z., and N.H. ensembled samples. F.C. and N.H. carried out the measurements. F.C., N.H., L.Z., Q.C., Y.H., M.T., Y.P., W.L., J.W., R.Z., Y.R., and X.L. analysed data. F.C., S.P., R.X.Z., and Y.Y. wrote the manuscript with input from H.C. and W.Y.. H.C. and Y.Y. supervised the project.

**Competing interests** The authors declare no competing interests.

**Correspondence and requests for materials** should be addressed to Wenyan Yin, Hongsheng Chen, Rui-Xing Zhang, andYihao Yang.